\documentclass[pra,twocolumn,tightenlines,nofootinbib]{revtex4}
\usepackage{bm,dcolumn,amsmath,graphicx,amsfonts,amssymb}
\usepackage{epsfig}
\usepackage{xcolor}

\begin{document}
\title{Nuclear deformation as a source of the non-linearity of King plot in the Yb$^+$ ion.}

\author{Saleh O.  Allehabi, V. A. Dzuba, and V. V. Flambaum}

\affiliation{School of Physics, University of New South Wales, Sydney 2052, Australia}

\author{A. V. Afanasjev}

\affiliation{Department of Physics and Astronomy, Mississippi State University, Mississippi 39762, USA}

\begin{abstract}
We perform atomic relativistic many-body calculations of the field isotope shifts and calculations of corresponding nuclear parameters for all stable even-even isotopes of Yb$^+$ ion.
We demonstrate that if we take nuclear parameters  of the Yb isotopes
from a range of   the state-the-art nuclear models which all predict strong quadrupole nuclear deformation, then calculated non-linearity of the King plot, caused by the difference in the deformation in different isotopes, is consistent with the non-linearity observed in the experiment (Ian Counts {\em et al},  Phys. Rev. Lett. {\bf 125}, 123002 (2020)). 
 The changes  of nuclear RMS radius between isotopes extracted from experiment 
are consistent with  those obtained in the  nuclear calculations.
\end{abstract}

\date{ \today }
\maketitle

In recent paper~\cite{Yb+IS}  the non-linearity of the King plot has been observed. The authors state that the effect may indicate physics 
beyond the Standard Model (SM), or, within the SM, may come from the quadratic field shift (QFS). Possible non-linearity of the King plot in 
Yb$^+$ was studied theoretically in Ref.~\cite{Mikami}. In the  present paper we show that it is more likely that the observed non-linearity of 
the King plot is due to a significant non-monotonic variation of the  nuclear deformation in the chain of isotopes.  We perform nuclear and 
atomic  calculations of the field isotope shift (FIS) which include nuclear  deformation and demonstrate that the dependence  of the 
deformation on isotopes  leads to a non-linearity of the King plot which is consistent with the observations in Ref. ~\cite{Yb+IS}.
We show that  the comparison of  theoretical and experimental  non-linearities  can be used to discriminate between different 
  nuclear models, favoring some and disfavoring others. 

It is well known from  experimental nuclear rotational spectra \cite{Eval-data}
%\cite{NuclearSpectra} 
 and its theoretical interpretation \cite{NilRag-book,ZHA.20} as well as from  presented below nuclear 
calculations  that all  even-even Yb isotopes  studied in \cite{Yb+IS} have deformed nuclear ground states
with the parameters of the quadrupole deformation $\beta \sim 0.3$.
In  our previous paper~\cite{prc} we demonstrated that nuclear deformation may lead to a non-linearity of the King plot.

Therefore,  in the present paper we calculate FIS in  even-even Yb isotopes  with accounting of
 nuclear deformation. We treat Yb$^+$ as a system with one external electron above closed 
shells and use the correlation potential method~\cite{CPM}.  We calculate the correlation potential $\hat \Sigma$  in 
the second order of the many-body perturbation  theory. Correlation potential is the non-local (integration) operator  
responsible for the correlation corrections due to interaction between valence electron and electrons in the core. 
Then we use $\hat \Sigma$ to calculate  the states of valence electron  (numerated by $v$)  in  the form of the Brueckner orbitals (BO)
\begin{equation}
\label{e:BO}
(\hat H^{\rm HF} + \hat \Sigma  -\epsilon_v)\psi_v^{\rm BO}=0.
\end{equation}
Here $\hat H^{\rm HF}$ is the relativistic Hartree-Fock (HF) Hamiltonian for the closed-shell core of Yb$^+$,
\begin{equation}
\label{e:HF}
\hat H^{\rm HF} = c\hat{\alpha_i}\cdot\hat{p_i} + (\beta -1)mc^2 + V_{\rm nuc}(r_i) + V_{\rm core}(r_i).
\end{equation}
In this expression  $\alpha$ and $\beta$ are  the Dirac  matrices, $V_{\rm nuc}$ is nuclear potential obtained by integrating nuclear charge density, $V_{\rm core}$ is the self-consistent HF potential  and the index $i$ numerates single-electron states.

FIS is calculated by varying nuclear potential $V_{\rm nuc}$  in (\ref{e:HF}).  
The RPA+BO method is similar to the MBPT (Many Body Perturbation Theory) method used in \cite{Yb+IS}.
 The results are presented in the form (see also \cite{Yb+IS}), in which index $a$ numerates atomic transitions,
\begin{equation} \label{e:FG}
\nu^{\rm FIS}_a = F_a\delta \langle r^2 \rangle  + G^{(2)}_a\delta \langle r^2 \rangle^2 + G^{(4)}_a\delta \langle r^4 \rangle. 
\end{equation}
First term in this equation is the standard FIS, other two terms are corrections responsible for the non-linearity of the King plot. 
The term with $ G^{(2)}_a$ is due to  the second order effect in the change of the nuclear Coulomb potential called  the quadratic field shift  (QFS)
 and the last term appears mainly due to the relativistic  effects in the electron wave function,  i.e. these terms represent different physical phenomena. 
On the other hand, their effects on the isotope shift are similar. It was suggested in Ref.~\cite{Yb+IS} that  $\langle r^4 \rangle$ and $\langle r^2 \rangle$ are related by $\langle r^4 \rangle = b \langle r^2 \rangle^2$, where $b$ is just a numerical constant, $b=1.32$. 
Extra care should be taken in calculating $G^{(2)}$ and $G^{(4)}$ independently on each other.
 For example,  they cannot be  defined  simultaneously in a fitting procedure. 
Therefore, we start the calculations  by eliminating the QFS term,  i.e. by considering FIS in the linear approximation. 
The change of the nuclear Coulomb  potential between two isotopes is considered as a perturbation and is treated in the first order using the random phase approximation (RPA). The RPA equations for core electrons have the following form \cite{CPM}:
\begin{equation}
\label{e:RPA}
(\hat H^{\rm HF}  -\epsilon_c)\delta \psi_c=-(\delta V_N + \delta V_{\rm core})\psi_c,
\end {equation}
where $\delta V_N$ is the difference between nuclear potentials for the two isotopes, index $c$ numerates states in the core, $\delta V_{\rm core}$ is the change of the self-consistent HF potential induced by $\delta V_N$ and  the  changes to all core functions $\delta \psi_c$. 
 The equations (\ref{e:RPA}) are solved self-consistently for all states in the core with the aim of finding  $\delta V_{\rm core}$.
The FIS for a valence state $v$ is then given by
\begin{equation}
\label{e:FIS}
\nu^{\rm FIS}_v = \langle \psi_v^{\rm BO} | \delta V_N + \delta V_{\rm core}| \psi_v^{\rm BO} \rangle.
\end{equation}
Apart from eliminating the QFS, an important advantage of using the RPA method (where the small parameter,  i.e. the change of the nuclear radius, is explicitly separated) 
%instead of studying the change of BO energies caused by the change of nuclear parameters 
is the suppression of a numerical noise. Non-linearity of the King plot is extremely  small and direct full scale calculations of the change of the atomic electron energy due to a tiny  change of the nuclear radius (i.e. without the separation of the small parameter) may  lead to a false effect in the King plot non-linearity (see below).  After FIS is calculated for a range of nuclear parameters, the constants $F_a$ and $G^{(4)}_a$ are found by fitting the results of the atomic calculations by formula (\ref{e:FG}) (without $G^{(2)}$) by the least-square-root method.

To calculate $G^{(2)}$ we use the second-order perturbation theory
\begin{equation} \label{e:e2}
G^{(2)}_a = \sum_n \frac{\langle a |\delta V_N + \delta V_{\rm core}| n \rangle^2}{E_a - E_n} / \delta \langle r^2 \rangle^2.
\end{equation}
Here $\delta V_N$ is the change of nuclear potential between two isotopes.
Summation goes over complete set of the single-electron basis states, including states in the core and negative-energy states. 
To include the core-valence correlations one can use BO for single-electron states $a$ and $n$. Again, the perturbation theory is used instead of the direct calculation of the change of the electron energy due to the tiny change of the nuclear radius to suppress numerical noice.

Instead of the direct summation over electron states in Eq. (\ref{e:e2}) one can  first solve
  the RPA equation for the valence state $a$ 
\begin{equation}
\label{e:RPAv}
(\hat H^{\rm HF}   + \hat \Sigma -\epsilon_a)\delta \psi^{\rm BO}_a=-(\delta V_N + \delta V_{\rm core})\psi^{\rm BO}_a,
\end {equation}
and then use
\begin{equation} \label{e:s2}
 G^{(2)}_a = \langle \delta \psi^{\rm BO}_a |\delta V_N + \delta V_{\rm core}| \psi^{\rm BO}_a \rangle / \delta \langle r^2 \rangle^2.
 \end{equation}
 We obtain the same results using Eqs.  (\ref{e:e2}) and (\ref{e:s2}). This provides a test of the numerical accuracy.

%*************

\paragraph{Nuclear deformation.}

  The quadrupole nuclear deformation $\beta$ provides a measure of the deviation of the nuclear density 
distribution from spherical shape so that nuclear radius $r_n(\theta)$ in the $\theta$ direction with respect 
of the axis of  symmetry is written as $r_n(\theta) = r_0(1+\beta Y_{20}(\theta))$.   
Electron feels nuclear density averaged over the nuclear  rotation (see e.g. Ref. ~\cite{prc}). 
We calculate the  average density by integrating the deformed density over $\theta$.

To determine the values of  $F$ and $G^{(4)}$ parameters in Eq. (\ref{e:FG}), we first
 vary the nuclear  root-mean-square (RMS) charge radius $r_c$ and the quadrupole deformation 
parameter $\beta$ in the  range determined by the nuclear theory (see below): 
 $5.234 \ {\rm fm} \leq r_c \leq 5.344 \ {\rm fm}$ and  $0.305 \leq \beta \leq 0.345$,  and 
 then fit the $F$ and $G^{(4)}$ parameters  by the formula (see also \cite{e120,prc}) 
\begin{eqnarray}
%&&\delta \nu^{\rm FS} = F\delta \langle r^2 \rangle +  d \Delta \beta, \label{e:Fd} \\
&&\nu^{\rm FIS} = F\delta \langle r^2 \rangle  +  G^{(4)}\delta \langle r^4 \rangle. \label{e:FG1}
\end{eqnarray}
 to the results of atomic calculations of FIS for different  $r_c$ and $\beta$.
The values of  $F$ and $G^{(4)}$ parameters  defined in such a way
are presented in Table~\ref{t:fit}. 
The table also  gives  the values of the $G^{(2)}$ parameters calculated using (\ref{e:e2}) and (\ref{e:s2}). Note that  FIS for  the $d$ states of Yb$^{+}$ is about 2 orders of magnitude smaller than FIS for the $6s$ states and  in QFS small matrix elements for  the $d$ states appears in  the second-order while in the  calculations of $F$ in the first order.  Therefore, the relative difference in  the $G^{(2)}$ parameters  for the 
 $s-d_{3/2}$ and $s-d_{5/2}$ transitions  is much smaller than the relative difference  for the $F$ parameters.

It was shown in Ref.~\cite{Yb+IS} that $\langle r^4 \rangle \approx b \langle r^2 \rangle^2$, where $b$ is just a numerical constant, b=1.32~\cite{Yb+IS}. 
We found that the situation is different in deformed and spherical  nuclei.
By calculating $\langle r^4 \rangle $ in both cases we found that the results can be fitted with high accuracy by the formula
\begin{equation}
\label{e:b}
\langle r^4 \rangle =\left[b_0 + b_1(r_c^2-r_0^2) + b_2(\beta-\beta_0)  \right] r_c^2, 
\end{equation}
where $r_0 = 5.179~{\rm fm}$ and  $\beta_0=0.305$.
For deformed nuclei $b_0=1.3129$, $b_1=-0.0036$, $b_2=0.1$, while for  spherical
 nuclei $b_0=1.2940$, $b_1=-0.0038$, $b_2=0$.
%Using (\ref{e:b}) one can make FIS explicitly dependent on the nuclear deformation. Such form of FIS was used in our previous paper~\cite{e120}.

\begin{table}
  \caption{Calculated parameters of formula (\ref{e:FG}) for the FIS in two transitions of Yb$^+$;
  $a$ stands for the $6s_{1/2} - 5d_{5/2}$  transition and $b$ stands for the $6s_{1/2} - 5d_{3/2}$ transition.
  Case 1 corresponds to  deformed nuclei, while case 2 corresponds to  spherical  nuclei.}
  \label{t:fit}
\begin{ruledtabular}
  \begin{tabular}{cc ddd}
\multicolumn{1}{c}{Case} &
\multicolumn{1}{c}{Tran-} &
\multicolumn{1}{c}{$F$} &
\multicolumn{1}{c}{$G^{(2)}$} &
\multicolumn{1}{c}{$G^{(4)}$} \\
&
\multicolumn{1}{c}{sition} &
\multicolumn{1}{c}{GHz/fm$^2$} &
\multicolumn{1}{c}{GHz/fm$^4$} &
\multicolumn{1}{c}{GHz/fm$^4$} \\
    \hline
%1 & $a$  & -16.5099 &   0.0 &    0.8684 & $10^{-3}$ \\ 
 %  & $b$  & -16.8768 &   0.0  &   0.8867 & $10^{-3}$ \\  
%2 & $a$  & -17.3653 &  0.01226 & 0.0 & $6\times10^{-5}$ \\ 
%   & $b$  & -17.7505 &  0.01252 & 0.0  & $6\times10^{-5}$ \\  

1 &  $a$  &-17.6035 &  0.02853 & 0.01308 \\
   &  $b$  &-18.0028  & 0.02853 & 0.01337 \\

2 & $a$  & -18.3026 &  0.02853 & 0.01245 \\
   & $b$  & -18.7201 &  0.02853 & 0.01273 \\

\end{tabular}
\end{ruledtabular}
\end{table}

\begin{table*}
  \caption{Nuclear RMS  charge radii ($r_c$, fm) and  the parameters of quadrupole deformation ($\beta$)  of even-even Yb isotopes 
 obtained  in different nuclear models.  The results obtained in the CDFT are labeled by the names of respective functionals.}
  \label{t:nuc}
\begin{ruledtabular}
  \begin{tabular}{lcccccccccccc}
&\multicolumn{2}{c}{BETA} &
\multicolumn{2}{c}{NL3*} &
\multicolumn{2}{c}{DD-ME2} &
\multicolumn{2}{c}{DD-ME$\delta$} &
\multicolumn{2}{c}{DDPC1} &
\multicolumn{2}{c}{FIT} \\
\multicolumn{1}{c}{$A$} &
\multicolumn{1}{c}{$r_c$} &
\multicolumn{1}{c}{$\beta$} &
\multicolumn{1}{c}{$r_c$} &
\multicolumn{1}{c}{$\beta$} &
\multicolumn{1}{c}{$r_c$} &
\multicolumn{1}{c}{$\beta$} &
\multicolumn{1}{c}{$r_c$} &
\multicolumn{1}{c}{$\beta$} &
\multicolumn{1}{c}{$r_c$} &
\multicolumn{1}{c}{$\beta$} &
\multicolumn{1}{c}{ $r_c$} &
\multicolumn{1}{c}{ $\beta$} \\
\hline
%             beta                        Nl3*                    DD-ME2                      DD-MEdelta                  DDPC1                        FIT
168 & 5.2950 & 0.3220 & 5.28751 & 0.33186 & 5.29144 & 0.33115 & 5.28820 & 0.33400 & 5.29528 & 0.33790 & 5.2950  &  0.3100 \\
170 & 5.3081 & 0.3258 & 5.30500 & 0.33873 & 5.31000 & 0.34028 & 5.30106 & 0.33070 & 5.31318 & 0.34540 & 5.3081  &  0.3184 \\
172 & 5.3204 & 0.3302 & 5.31678 & 0.33188 & 5.32056 & 0.33124 & 5.31138 & 0.32024 & 5.32346 & 0.33429 & 5.3204  & 0.3232 \\
174 & 5.3300 & 0.3249 & 5.32776 & 0.32203 & 5.33066 & 0.32072 & 5.32356 & 0.31447 & 5.33291 & 0.32152 & 5.3300  & 0.3208 \\
176 & 5.3391 & 0.3050 & 5.33868 & 0.31471 & 5.34230 & 0.31436 & 5.33228 & 0.30413 & 5.34420 & 0.31398 & 5.3391  &  0.3156 \\
\end{tabular}
\end{ruledtabular}
\end{table*}

To study the non-linearity of the King plot we need total isotope shift (including mass shift) for two transitions $a$ and $b$.
%(QUESTIONS/SUGGESTION. SHOULD WE WRITE "for two transitions a and b" HERE? THIS WILL
%ALSO EXPLAIN THE MEANING OF SUBSCRIPTS "a" and "b" IN THE FORMULA BELOW.)
Then using Eq.  (\ref{e:FG}) one can write for the isotope shift between isotopes $i$ and $j$
\begin{eqnarray}
%\label{e:king}
&&\frac{\nu_{bij}}{\mu_{ij}} = \frac{F_b}{F_a}\frac{\nu_{aij}}{\mu_{ij}} + \left(K_b - \frac{F_b}{F_a}K_a\right) + \nonumber \\
&&+ \left(G^{(2)}_b - \frac{F_b}{F_a}G^{(2)}_a\right)\frac{\delta \langle r^2 \rangle_{ij}^2}{\mu_{ij}} + \label{e:king} \\
&&+ \left(G^{(4)}_b - \frac{F_b}{F_a}G^{(4)}_a\right)\frac{\delta \langle r^4 \rangle_{ij}}{\mu_{ij}}. \nonumber
%&& + \left(d_b - \frac{F_b}{F_a}d_a\right) \frac{\Delta \beta_{ij}}{\mu_{ij}}. \nonumber
\end{eqnarray}
%Similar expression can be obtained for Eq. (\ref{e:FG}).
Here $K$ is the electron structure factor for the mass shift, $\mu = 1/m_i - 1/m_j$ is the inverse mass difference.
First line of Eq. (\ref{e:king}) corresponds to the standard King plot, second and third lines contain the terms which may cause the King plot non-linearities.

To study these non-linearities we  use the least-square fitting of Eq. (\ref{e:king}) by the formula $\nu_b' = A\nu_a' + B$, where $\nu'=\nu/\mu$.
The relative non-linearities are calculated as $\Delta \nu_b'/\nu_b'$, where $\Delta \nu_b'$ is the deviation of the isotope shift $\nu'_b$ from its linear fit.
To do the fitting and making King plot we need to know the change of nuclear parameters $\delta \langle r^2 \rangle$ and $\Delta \beta$ between the isotopes of interest. We use nuclear calculations for this purpose.  Nuclear parameters of the Yb isotopes with even neutron number
obtained in different  nuclear  models  are presented in Table~\ref{t:nuc}.

%The simplest model,  called BETA, is based on the estimations of  the ratios of the nuclear deformation parameters $\beta$ for different isotopes 
%extracted from experimental nuclear spectra assuming  $\beta^2 \propto 1/\omega$, where $\omega$ is the frequency of the nuclear electric quadrupole %transition between  the ground and first  rotational  states\footnote{The energy interval between rotational levels is inversely proportional to the 
%moment of inertia $J$  and in a fluid nucleus  model in book  \cite{nucstr} the moment of inertia  $J \propto \beta^2$ (see also 
%Ref.  \cite{Minkov2017} where the corresponding energy  term has $\beta^2$ in the denominator).}.

%{\color{red}  In the simplest model, called BETA, the nuclear deformation parameter $\beta$  in a 
%given nucleus is extracted from experimental energy $E_{2^+}$ of the first rotational state with spin 
%$I=2^+$ in the ground state rotational band  via Grodzins relations \cite{NilRag-book}
%
%\begin{equation}
%\beta = \sqrt{\frac{1225}{A^{7/3} E_{2^+}}}
%\end{equation}
%
%where $A$ stands for mass number of the nucleus and $E_{2^+}$ is given in MeV. }
In the simplest model, called BETA, the nuclear deformation parameter $\beta$  in a 
given nucleus is extracted from measured reduced electric quadrupole transition rate B(E2) for the ground state to $2^+$ state transition.
These values are tabulated  in Ref. \cite{RNT.01}. 
We also introduce a hypothetical model (labeled as FIT) which has nuclear parameters leading to very accurate fit of both experimental FIS 
and  the deviations of King plot from linearity.
Note that the parameters of the FIT model are not so different from other models, i.e. they are pretty realistic.

    The ground state properties  of  the nuclei under study have also been calculated 
within the Covariant Density Functional Theory (CDFT) using several 
state-of-the-art covariant energy density functionals (CEDFs)
such as DD-ME2, DD-ME$\delta$, NL3* and DD-PC1 \cite{AARR.14}. In the CDFT, the nucleus is considered as a system of 
$A$ nucleons which interact via the exchange of different mesons and nuclear many-body
correlations are taken into account. Above mentioned CEDFs represent 
three major classes of covariant density functional models which provide accurate description
of the ground state properties (such as deformations, charge radii, etc.) of even-even nuclei  across
the nuclear chart \cite{AARR.14,AA.16}. The  main differences between them
lie in the treatment of the interaction range and density dependence. The best global description of experimental
data on charge radii has been achieved by the DD-ME2 functional [characterized by 
RMS deviation of $\Delta r^{rms}_{ch}=0.0230$ fm], followed by DD-PC1 [which also provides best 
global description of binding energies], NL3* and finally by DD-ME$\delta$ 
[characterized by RMS deviation of $\Delta r^{rms}_{ch}=0.0329$ fm] (see Table VI in Ref.\ \cite{AARR.14}
and Fig. 7 in Ref.\ \cite{AA.16}).

%The accuracy of the description of the ground state properties (such as 
%binding energies, charge radii etc) of even-even nuclei has been investigated 
%globally in Refs.\ \cite{AARR.14,AA.16}. The best global description of experimental
%data on charge radii has been achieved by the DD-ME2 functional [characterized by 
%rms deviation of $\Delta r^{rms}_{ch}=0.0230$ fm], followed by DD-PC1 [which also provides best 
%global description of binding energies], NL3* and finally by DD-ME$\delta$ 
%[characterized by rms deviation of $\Delta r^{rms}_{ch}=0.0329$ fm] (see Table VI in Ref.\ \cite{AARR.14}
%and Fig. 7 in Ref.\ \cite{AA.16}). However, the spread of rms deviations for charge 
%radii between above  mentioned functionals is rather small  ($\Delta (\Delta r^{rms}_{ch}) = 0.0099$ fm). 
%On the other hand, the charge radii of some isotopic chains (especially, those with high proton 
%number $Z$) are not very accurately measured. Thus, strictly speaking we have to consider the 
%accuracy of the description of charge radii by these functionals as comparable.

\begin{figure}[tb]
\epsfig{figure=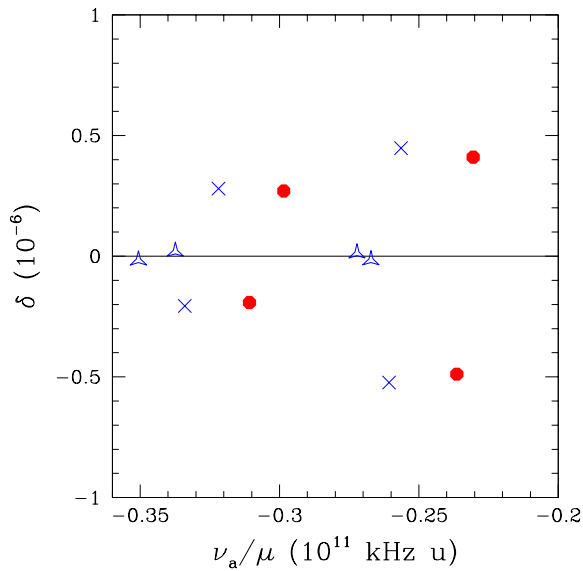,scale=0.4}
\caption{ The deviations from linear King plot in experiment (solid red circles) and theory.
Theoretical deviations caused by nuclear deformation are shown as blue crosses,  and those by QFS are shown as blue triangles.
All theoretical numbers correspond to the FIT nuclear model.}
% Note the two times difference in vertical scale with Fig.~\ref{f:nld}.}
\label{f:nl2}
\end{figure}

Using the parameters coming from these models  we calculate FIS, build the King plot, find its deviations from the linearity and compare the results to the experimental data from Ref.  ~\cite{Yb+IS}. 
The results are presented in Table~\ref{t:models} and Fig.~\ref{f:nl2}.
One can see that the values of the experimental and theoretical non-linearities are of the same order of magnitude for all nuclear models. 
This already means that the nuclear deformation is an important effect which has to be included into the analysis. 
Moreover, for some models (e.g., BETA, FIT, NL3*, DDPC1) there is a strong correlation between experimental and theoretical data. 

To make sure that the non-linearities come from the nuclear deformation and not from QFS, we perform two tests. 
In the first test we remove nuclear deformation from the calculations by  using the values of $\langle r^4 \rangle$ in (\ref{e:FG}) which come from the calculations assuming that all isotopes have  spherical shapes.
 In the second test  we put $b_2=0$ in Eq. (\ref{e:b}). In both cases the deviations of the King plot from the linearity  drop by about an order of magnitude.
This means that the nuclear deformation is likely to be the main source of the observed non-linearity of the King plot.

%To make sure that the non-linearity comes from nuclear deformation and not QFS we performed calculations for a model case in which the parameter $\beta$ of nuclear deformation was kept the same for all isotopes ($\beta=0$ and $\beta=0.3$). The results are fitted by first two terms in (\ref{e:fit}) to the accuracy of $10^{-5}$ in both cases. The resulting non-linearity of King plot $\sim 10^{-9}$, i.e. three orders of magnitude smaller that in the experiment.

%To summarise the arguments for the nuclear deformation being the most likely source of the observed non-linearity of King plot we state that
%(a) calculations with the same nuclear deformation or no nuclear deformation for all isotopes produce no non-linearities; (b) calculations with quadrupole nuclear deformation taken from a range of nuclear models produce non-linearities consistent with the observations in all cases; (c) accurate fitting of both, isotope shift and non-linearities is achieved with realistic nuclear parameters; (d) the only other important source of the non-linearity within standard model, the QFS, is about an order of magnitude smaller then the effect of nuclear deformation; (e) QFS cannot explain non-regular deviations from the linear plot observed in experiment, while nuclear deformation can.

%\begin{figure}[tb]
%\epsfig{figure=nld.eps,scale=0.4}
%\caption{Deviations from linear King plot in experiment (solid red circles) and theory. Theoretical results include nuclear deformation and
%use different nuclear models indicated in right top conner.}
%\label{f:nld}
%\end{figure}

\begin{table*}
  \caption{The deviations from the linearity of the King plot (in parts of $10^{-6}$).  The comparison between experiment~\cite{Yb+IS} and calculations in different nuclear models} %$r$ is  the Pearson correlation coefficient between theoretical and experimental data.}
  \label{t:models}
\begin{ruledtabular}
  \begin{tabular}{c ddddddd}
\multicolumn{1}{c}{Isotope} &&
\multicolumn{6}{c}{Nuclear model} \\
\multicolumn{1}{c}{pair} &
\multicolumn{1}{c}{Expt.} &
\multicolumn{1}{c}{BETA} &
\multicolumn{1}{c}{FIT} &
\multicolumn{1}{c}{NL3*} &
\multicolumn{1}{c}{DD-ME2} &
\multicolumn{1}{c}{DD-ME$\delta$} &
\multicolumn{1}{c}{DDPC1} \\
    \hline
%\multicolumn{8}{c}{using formula (\ref{e:Fd})} \\
%168 - 170 & -0.192 &  0.642  & -0.233 & -0.492 & -0.003 &  -0.304 &  -0.018 \\
%170 - 172 &  0.270 &  -1.24 &  0.320 &  0.408 &  0.115 &  0.514 &  -0.425  \\
%172 - 174 & -0.489 &  -3.07 & -0.457 & -1.42  &  -0.084 &  0.459 & 0.356 \\
%174 - 176 &  0.411  &   3.17 &  0.370 &  1.41  &  -0.029 & -0.685 &  0.079 \\
%\hline
%           &    $r$    & 0.6595 &  0.9934 & 0.9780 & 0.5802 & -0.4214 & -0.6279 \\
%    \hline
%\multicolumn{8}{c}{using formula (\ref{e:FG})} \\
168 - 170 & -0.192 &  0.642  & -0.206 & -0.037 & -0.084 &  -0.511 &  -0.080 \\
170 - 172 &  0.270 &  -0.607 &  0.281 & -0.159 &  -0.467 &  0.546 &  -0.222  \\
172 - 174 & -0.489 &  -3.05 & -0.523 & -0.200  &  -0.028 &  0.392 & -0.198 \\
174 - 176 &  0.411  &   3.03 &  0.448 &  0.387  &   0.551 & -0.406 &  0.472 \\
%\hline
       %    &    $r$    & 0.6975 &  0.9999 & 0.6601 & 0.2480 & -0.1454 & 0.6071 \\
\end{tabular}
\end{ruledtabular}
\end{table*}

\paragraph{Quadratic field  shift.}

\begin{table}
  \caption{ The deviations from the linearity of the King plot $\delta$ due to the quadratic field shift.  The comparison between experiment~\cite{Yb+IS} and calculations using  the $\delta \langle r^2 \rangle$  values which fit the experimental isotope shift ~\cite{Yb+IS}. The deviation $\delta$ is shown as a function of $\nu_a/\mu$ (see Eq.~(\ref{e:king})).}
    \label{t:QFS}
\begin{ruledtabular}
  \begin{tabular}{c dddd}
\multicolumn{1}{c}{Isotope} &
\multicolumn{2}{c}{Expt.} &
\multicolumn{2}{c}{QFS} \\
\multicolumn{1}{c}{pair} &
\multicolumn{1}{c}{$\nu_a/\mu$} &
\multicolumn{1}{c}{$\delta$} &
\multicolumn{1}{c}{$\nu_a/\mu$} &
\multicolumn{1}{c}{$\delta$} \\
&\multicolumn{1}{c}{$10^{11}$ kHz u} & 
\multicolumn{1}{c}{$10^{-6}$} & 
\multicolumn{1}{c}{$10^{11}$ kHz u} & 
\multicolumn{1}{c}{$10^{-6}$}  \\
    \hline
%168 - 170 & -0.311 & -0.192 & -0.330 & -0.007 \\
%170 - 172 & -0.299 &  0.270 & -0.318 & 0.007  \\
%172 - 174 & -0.236 & -0.489 &  -0.254 &  0.024 \\
%174 - 176 & -0.231  &  0.411 &  -0.247 &  -0.024 \\
168 - 170 & -0.311 & -0.192 & -0.351 & -0.017 \\
170 - 172 & -0.299 &  0.270 & -0.337 & 0.020  \\
172 - 174 & -0.236 & -0.489 &  -0.272 &  0.013 \\
174 - 176 & -0.231  &  0.411 &  -0.267 &  -0.016 \\
\end{tabular}
\end{ruledtabular}
\end{table}
%  nu/mu         dv             nu/mu       dv
%-0.3108E+11 -0.1921E-06    -0.3300E+11 -0.2755E-07  
%-0.2985E+11  0.2700E-06    -0.3179E+11  0.3417E-07 
%-0.2364E+11 -0.4886E-06    -0.2544E+11 -0.8120E-08 
%-0.2305E+11  0.4106E-06    -0.2470E+11  0.1198E-08 

Ref.~\cite{Yb+IS} argues that QFS is the main source of  the non-linearity of the King plot. However, their calculations only provided an upper limit on the non-linearity since the results of CI and MBPT calculations were very different.  From our point of view the problem with the calculations in  Ref.~\cite{Yb+IS} is that they have not separated a small parameter, the change of the nuclear radius, and obtained FIS from the small difference in 
the energies of the atomic transitions  
calculated for different nuclear radii.
This is certainly a good approach for the calculation of FIS but it is not good enough to calculate a very small non-linearity which is extremely sensitive to  numerical noice.  
 
Our results presented earlier 
%{\it in the previous section COMMENT: WE DO NOT HAVE SECTIONS} 
indicated that QFS gives a much smaller contribution to the non-linearity of the King plot than the upper limit presented in Ref.~\cite{Yb+IS}.  To test this result we performed  FIS and QFS calculations by a different method
%However, it is an important effect which was claimed as the main source of the non-linearity in Ref.~\cite{Yb+IS}. Therefore, it deserves a more detailed consideration.
assuming that all isotopes have spherical nuclear shape
%spherically-symmetric nuclei 
($\beta=0$). 
%As it was mentioned above,  RPA method cannot be used to find QFS  because it is linear in external perturbation.
The main motivation for using  RPA method in the case of nuclear deformation is the minimization of 
numerical noice which comes from extra integration over directions. 	
There is no such problem for spherical  nuclei and the procedure is less complicated. FIS in this case may be found from the direct variation of the nuclear radius in the nuclear Coulomb potential.  We perform HF and BO calculations for a range of nuclear  charge
RMS radii from $\langle r^2 \rangle$=(5~fm)$^2$ to $\langle r^2 \rangle$=(6~fm)$^2$ and present the results by the same formula (\ref{e:FG}) (see Table~\ref{t:fit}). 
%WHERE THESE RESULTS ARE PRESENTED? 
As in case of deformed nuclei, the QFS parameter $G^{(2)}$ is found from the perturbation theory calculations. 
The values of $F$ and $G^{(4)}$ are slightly different.

The same equation (\ref{e:king}) and  the same procedure were used to find the non-linearities of the King plot.
The results are presented on Fig.~\ref{f:nl2} and Table ~\ref{t:QFS}. As one can see, the non-linearity caused by QFS is an order of magnitude smaller than the observations. It is also much smaller than the non-linearity caused by the variation of the nuclear deformation. 

We also  performed another test calculation using constant value $\beta=0.3$ instead of $\beta$=0. Again, without variation of   $\beta$ the non-linearity of the King plot is small.

\paragraph {The change of nuclear  RMS  charge radius.}

Formula (\ref{e:FG}) with parameters $F, G^{(2)},  G^{(4)}$ from Table~\ref{t:fit}  can be used to find the change of the nuclear RMS charge radius between isotopes by fitting experimental FIS. 
The values of the $\delta \langle r^2 \rangle$ corresponding to the best fit (the FIT model in Table~\ref{t:nuc}) are presented in Table~\ref{t:dr} and compared with other data. 
Note that if  FIS is calculated using parameters of other nuclear models from Table~\ref{t:nuc} then 
 the difference between theory and experiment ranges from few  percent to $\sim$~15\%. 
 This is because nuclear theory is not  sufficiently accurate in predicting $\delta \langle r^2 \rangle$.
  It is easy to see from the data in Table~\ref{t:nuc} that 0.01\% change in the nuclear RMS radius may lead to $\sim$~10\% change in the value of $\delta \langle r^2 \rangle$ leading to the same change in FIS. Note, however, very good agreement for the $\delta \langle r^2 \rangle$ between best fit and the predictions of the DD-ME$\delta$ nuclear model (see Tables \ref{t:nuc} and \ref{t:dr}). This might be fortuitous.
This model is not the best in reproducing experimental non-linearities of King plot. 
We stress that the non-linearities of King plot are more sensitive to the change of nuclear shape  rather than to its  RMS charge radius.
% We stress that {\bf for this functional ??} the change of the nuclear RMS radius fits the FIS but has little effect on the non-linearities of the King plot. And vice versa, nuclear shape fits the non-linearities of the King plot while having little effect on the total value of FIS if correct value of $\delta \langle r^2 \rangle$ is used (to avoid misunderstanding note that $\beta$ significantly affects $\langle r^2 \rangle$) .  
 
\begin{table}
  \caption{The changes of nuclear RMS  charge radius ($\delta \langle r^2 \rangle$, fm$^2$) extracted from the isotope shift measurements.}
 \label{t:dr}
\begin{ruledtabular}
  \begin{tabular}{c cddd}
    \multicolumn{1}{c}{Isotope} &
    \multicolumn{2}{c}{Ref.~\cite{Yb+IS}} &
    \multicolumn{1}{c}{Ref.~\cite{Marinova,dr2}} &
    \multicolumn{1}{c}{This work} \\
    \multicolumn{1}{c}{pairs} &
    \multicolumn{1}{c}{CI} &
    \multicolumn{1}{c}{MBPT} & \\
    \hline
    (168,170) & 0.156 & 0.149   & 0.1561(3) & 0.138  \\
    (170,172) & 0.146 & 0.140   & 0.1479(1) & 0.130  \\
    (172,174)  & 0.115 & 0.110   & 0.1207(1) & 0.102 \\
    (174,176)  & 0.110 & 0.105   & 0.1159(1)  & 0.097 \\
\end{tabular}
\end{ruledtabular}
\end{table}

\paragraph{The comparison with other results for  $\delta \langle r^2 \rangle$.} It is instructive to analyze possible reasons for the difference between our 
results and other results for  $\delta \langle r^2 \rangle$  presented in Table~\ref{t:dr}.
There is a 12 to 19\% difference between our results and those  published in Ref.~\cite{Marinova} (see Table~\ref{t:dr}). However,  the latter
 were taken from a fifty-years-old paper~\cite{dr2} which has no many-body calculations but only estimations based on the single-electron consideration. The uncertainty of such estimations can be well above 10\% and even 20\%.
 %the difference {\color{red}  in $\delta \langle r^2 \rangle$ ???}.

There is also  a 8\% to 13\% difference between our results and those of Ref.~\cite{Yb+IS}.  
Ref.~\cite{Yb+IS} contains two calculations of the FIS constants performed by CI and MBPT methods with the 4\% difference between corresponding results.
Our FIS constant $F$ is about 13\% larger than the same constant calculated in Ref.~\cite{Yb+IS} using the CI method and about 8\% larger than those calculated in Ref.~\cite{Yb+IS} using the MBPT method. 
This explains the difference in the results for $\delta \langle r^2 \rangle$ (Table~\ref{t:dr}). When we use the numbers from Ref.~\cite{Yb+IS} in Eq.~(\ref{e:FG}) we reproduce their results for $\delta \langle r^2 \rangle$. The difference in the results seems to be due to the difference in the procedures defining the constants $F$ and $G$.  
We use BO and the RPA method to calculate $F$ and $G^{(4)}$ and the perturbation theory to find $G^{(2)}$ as it has been explained above.
The authors of Ref.~\cite{Yb+IS} calculate $F$ as a leading term of the Seltzer moment expansion at the origin for the total electron density (see Eq. (S11) in \cite{Yb+IS}) and then use  partial derivatives of FIS to calculate constants $G$. Such method looks sensitive to the degeneracy of $G^{(2)}$ and $G^{(4)}$ contributions to FIS.  An indication of the problem may be a significant relative difference in $G^{(2)}$ parameters in Ref.~\cite{Yb+IS} while we argued above  that it must be very small since it appears in the second order of the small $d$ wave  FIS matrix elements.

It is instructive to explain why the  ratios $G^{(4)} / F$ are different in the $s-d_{3/2}$ and $s-d_{5/2}$ transitions (this is needed for  the non-linearity of the King plot without QFS). We suggest the following mechanism supported by the numerical calculations.  According to it
only two relativistic Dirac wavefunctions, $s_{1/2}$ and $p_{1/2}$, penetrate into the nucleus. They  have different  spatial distributions inside  and therefore the ratios of  the $\delta \langle r^2 \rangle$ and $\delta \langle r^4 \rangle$ contributions to their energies and wavefunctions are  noticeably different.  The   $d_{3/2}$ and $d_{5/2}$ wavefunctions 
interact differently with the $s_{1/2}$ and $p_{1/2}$  ones and  this gives the difference in   $G^{(4)}/F$.

In conclusion we state that presented arguments indicate that nuclear deformation is the most likely source of recently observed non-linearities of King plot in Yb$^+$. The results of the combined nuclear and atomic calculations for the effect are consistent with the observations. The contribution of the QFS is about an order of magnitude smaller. The measurements of the non-linearity of the King may be used to study nuclear deformation in nuclei with zero spin where nuclear electric quadrupole moment can not be extracted from atomic spectroscopy.  The changes of nuclear charge  RMS radii between even-even Yb isotopes extracted from atomic measurements are consistent with nuclear theory.

We are grateful J. Berengut for useful discussion. The work was supported by the Australian Research Council grants No. DP190100974 and DP200100150 and  by
%. This material is based upon work supported  by
 the US Department of Energy, Office of Science, Office of Nuclear Physics under Grant No. DE-SC0013037

\end{document}